\documentclass[mathleft]{an}
\usepackage{graphicx}
\usepackage{times}

\begin{document}

\Pagespan{1}{}
\Yearpublication{2006}%
\Yearsubmission{2006}%
\Month{11}%
\Volume{999}%
\Issue{88}%

\def\gsimeq
{\hbox{\raise0.5ex\hbox{$>\lower1.06ex\hbox{$\kern-1.07em{\sim}$}$}}}
\def\lsimeq
{\hbox{\raise0.5ex\hbox{$<\lower1.06ex\hbox{$\kern-1.07em{\sim}$}$}}}
\def\pn{\par\noindent}
\def\ss{\smallskip\pn}
\def\ms{\medskip\pn}
\def\bs{\bigskip\pn}
\def\aap{A\&A}
\def\mnras{MNRAS}
\def\apj{ApJ}
\def\apjl{ApJ}
\def\nat{Nature}

\title{Relativistic blue- and red-shifted absorption lines in AGNs}

\author{M. Cappi\thanks{Corresponding author:
  \email{cappi@iasfbo.inaf.it}}
}
\titlerunning{Outflows and inflows in AGNs, a review}
\authorrunning{M. Cappi}
\institute{INAF-IASF Bologna, Via Gobetti 101, I-40129, Bologna, Italy}

\received{21 August 2006}
\accepted{15 September 2006}
\publonline{later}


\abstract{
Current, accumulating evidence for (mildly) relativistic blue- and 
red-shifted absorption lines in AGNs is reviewed. {\it XMM-Newton} and {\it Chandra} 
sensitive X-ray observations are starting to probe not only
the  kinematics (velocity) but also the dynamics (accelerations) of
highly ionized gas flowing in-and-out from, likely, a few gravitational radii
from the black hole. It is thus emphasized that X-ray absorption-line spectroscopy 
provides new potential to map the accretion flows near black holes, to 
probe the launching regions of relativistic jets/outflows, and to quantify the 
cosmological feedback of AGNs. Prospects to tackle these issues with future high
energy missions are briefly addressed.
}

\maketitle

\section{Introduction}

Little is known about the flow patterns of gas in the innermost regions of black holes 
(BHs) in AGNs. Probing of the gas kinematics (velocities) and, most importantly, dynamics (accelerations) 
around black holes is a fundamental requisite if we want to understand the source's geometry 
and energy generation mechanism, i.e. the accretion mode.

Surprisingly enough, there is little {\it direct} evidence yet for inward radial motions 
(e.g. infalling/inflowing material) down to a few gravitational radii where most of the 
energy is generated. Probably the most convincing evidence of accretion onto a relativistic 
black hole are, to date, the detections of broadened, {\it redshifted} FeK {\it emission} line in the 
X-ray spectra of an increasing number of bright Seyfert 1 galaxies (Tanaka et al. 1995, Fabian 2002, 
see contributions in these proceedings by Fabian et al., Guainazzi et al., and Nandra et al.). 
Nandra et al. (1999) discussed the detection of a redshifted 
Fe resonant {\it absorption} feature in the X-ray spectrum of the Seyfert 1 galaxy 
NGC3516. The $ASCA$ data showed a sharp and narrow drop of counts at E$\sim$5.9 keV 
(corresponding, for FeXXVI, to an energy/velocity shift of $\sim$0.15 c) 
imprinted over a broad emission line profile, with indication of variability 
on time scales of $\sim$20 ks. Being the line redshifted, the authors speculated its 
association with material free-falling onto the BH. Unfortunately, these early studies were
strongly limited by the instrumental sensitivity in the 2-10 keV band. 
The low statistic combined with a complex underlying continuum thus prevented any detailed studies. 

Several are, on the contrary, the evidences in AGNs of matter flowing outward. 
Fast winds/outflows/ejecta are seen in AGNs since a long time, i.e.
the jets of radio-loud galaxies (Axon et al. 1989, Tadhunter 1991), the (absorbed) spectra of 
broad absorption line (BAL) QSOs 
(Weymann et al. 1991, Reichard et al. 2003) or the [OIII] ionization cones of Seyfert galaxies 
(Tadhunter \& Tsvetanov 1989, Wilson \& Tsvetanov 1994). Warm 
absorbers are also common among AGNs. Indeed, they have been detected in more than half of bright 
Seyfert galaxies (Reynolds 1997, George et al. 1997, Blustin et al. 2005). 
High resolution UV and soft X-ray observations have shown that warm absorbers have typical 
temperatures of $\sim$10$^{5-6}$ K and outflowing velocities of a few 100 to a few 1000 km s$^{-1}$ 
(see Crenshaw, Kraemer \& George, 2003 for a comprehensive review on AGN winds and 
warm absorbers). 

With the superior sensitivity of {\it XMM-Newton} and {\it Chandra} between 2-10 keV, it has now been 
possible to probe these phenomena in more detail. In particular, more extreme 
cases of redshifted and blueshifted absorption structures have now been discovered 
in the spectra of a number of bright AGNs. This paper briefly reviews the latest results 
on these new, extreme phenomena.
Specifically, recent data show strong evidence for outflowing gas with unprecedented kinematic 
and ionization properties (\S 2.1 and 2.2), and provide hints of redshifted 
absorbers that would indicate either infalling clouds or strong gravitational 
redshift (\S 2.3 and 2.4). Despite some remaining observational critical issues (Section 3), 
X-ray absorption spectroscopy with future observatories (Section 4) offers new potential to 
probe the dynamics of the accretion and ejection of material onto black holes.

\begin{table*}
\small
\caption{}
\begin{center}
\begin{tabular}{llllllll}
\multicolumn{8}{c}{\bf Blue-shifted Fe lines} \\
\hline
\multicolumn{1}{l}{Name} &
\multicolumn{1}{l}{Type} &
\multicolumn{1}{l}{z} &
\multicolumn{1}{l}{E$_{rest}$} &
\multicolumn{1}{l}{v/c} &
\multicolumn{1}{l}{log N$_{\rm H}$} &
\multicolumn{1}{l}{log $\xi$} &
\multicolumn{1}{l}{Reference} \\
\multicolumn{1}{l}{} &
\multicolumn{1}{l}{} &
\multicolumn{1}{l}{} &
\multicolumn{1}{l}{(keV)} &
\multicolumn{1}{l}{} &
\multicolumn{1}{l}{(cm$^{-2}$)} &
\multicolumn{1}{l}{(erg cm s$^{-1}$)} &
\multicolumn{1}{l}{} \\
\hline
NGC1365       & Sey1.8 & 0.0055& 6.7-7.2 (K$_{\alpha}$) & 0.016 & 23-23.7 & 3.7 & 1\\ 
              &        &       & 7.8-8.3 (K$_{\beta}$)  &             &         &     &                      \\ 
MCG-6-30-15   & Sey1.2 & 0.0077& 6.74 and 7        & 0.007 & 23.2 & 3.6 & 2 \\
MCG-5-23-16   & Sey1.9 & 0.0085& 7.7               & 0.1   & 22.9   & 3.6 & 3 \\
NGC3783       & Sey1   & 0.0097& 6.7               & 0.003 & 22.7 & 3   & 4, 5 \\
IC4329a       & Sey1   & 0.0160& 7.7               & 0.1   & 22.1 & 3.7 & 6\\
IRAS13197-1627& Sey1.8 & 0.0165& 7.5               & 0.11  & 23.7 & $\gsimeq$3& 7 \\
Mrk509        & Sey1   & 0.034 & 8.2               & 0.1-0.2&23.1& 3.5 & 8 \\
PG0844+349    & Sey1   & 0.064 & 8.7               & 0.2   & 23.6 & 3.7 & 9 \\
PG1211+143    & NLSey1 & 0.081 & 7.6               & 0.13  & 22.3 & 2.9 & 10, 11 but see 12 \\
PDS456        & RQ QSO & 0.184 & 7.6-9.3           & 0.15  & 23.7 & 2.5 & 13 \\
PG1115+080    & BAL QSO& 1.72  & 7.4 and 9.5 & 0.1 and 0.34& 23.4 and 23.8 &  $\gsimeq$3 & 14\\
APM08279+5255 & BAL QSO& 3.91  & 8.1 and 9.8 & 0.2 and 0.4 & 23 & $\gsimeq$3 & 15 but see 16\\
\hline
\multicolumn{8}{c}{} \\
\multicolumn{8}{c}{\bf Red-shifted Fe lines} \\
\hline
Mrk335        & Sey1   & 0.026 & 5.9               & 0.15  & 22.4 & 3.6 & 17 \\
Mrk509        & Sey1   & 0.034 & 5.4-5.5           & 0.21  & 22.8 & 3.5 & 8 \\
PG1211+143    & NLSey1 & 0.081 & 4.56 and 5.33     & 0.4 and 0.26 & 23.6 & 3.9 & 18 \\
Q0056-363     & RQ QSO & 0.162 & 5.34              & 0.23  & $\gsimeq$23 & $\gsimeq$3 &  19 \\
E1821+643     & RQ QSO & 0.297 & 6.2               & 0.1   & -    & -   & 20 \\
\hline
\end{tabular}
\end{center}
\par\noindent
Col. (1): galaxy name. 
Col. (2): galaxy type.
Col. (3): redshift
Col. (4): Line rest-frame energy in units of keV.
Col. (5): Velocity (in units of c) corresponding to the line shift, assuming absorption by H-like Fe. 
Col. (6): Log of the absorption column density in units of atoms cm$^{-2}$.
Col. (7): Log of the ionization parameter in units of erg cm s$^{-1}$. 
Col. (8): References - 1) Risaliti et al. 2005; 2) Young et al. 2005; 3) Braito, this conference; 
4) Kaspi et al. 2002; 5) Reeves et al. 2004; 6) Markowitz et al. 2006; 7) Dadina \& Cappi 2004; 
8) Dadina et al. 2005; 9) Pounds et al. 2003; 10) Pounds et al. 2003; 11) Pounds et al. 2006; 
12) Kaspi et al. 2006; 13) Reeves et al. 2003; 14) Chartas, Brandt \& Gallagher 2003; 
15) Chartas et al. 2002; 16) Hasinger, Schartel \& Komossa 2002; 17) Longinotti et al. 2006; 
18) Reeves et al. 2005; 19) Matt et al. 2005; 20) Yaqoob \& Serlemitsos 2005.
\normalsize
\end{table*}

\section{X-Ray Absorption Lines}

\subsection{Blue-shifted Lines}

From recent observations with {\it XMM-Newton} and {\it Chandra}, there is now overwhelming evidence
of absorption lines at rest-frame energies $\sim$7--10 keV in the spectra of several radio-quiet (RQ) AGNs and 
QSOs. They are found in both low-z and high-z sources (see Table 1). 
If interpreted as absorption lines by He-like and H-like Fe, 
these imply the existence of massive, high velocity (v$\sim$0.1-0.4 c) and highly ionized 
outflows in AGNs, with rather extreme values of absorption column densities ($\sim$10$^{23-24}$ cm$^{-2}$) 
and ionization parameters (log$\xi$ $\gsimeq$  3).

The mass outflow rates sensitively depend on the absorber's distance from the ionization 
continuum source, on the assumed volume filling factor, and other 
unknown values (such as the gas density for example), but can be comparable or even larger than the 
Eddington accretion rates.

Most interestingly, variability on timescales down to few 1000 s has been found among the 
brighter sources, two remarkable examples of which are shown in Figure 1 and 2. A measure of how 
these ``hot'' absorbers respond to flux variations in time allows to place some constraints 
on their geometrical properties and their origin (e.g. Risaliti et al. 2005, Dasgupta et al. 2005).
Fitting with time-dependent photoionization models such as that calculated by Nicastro et al. (1999) 
will be a challenge for future observations.

These features are also reminiscent of the peculiar absorption structures found near 
1 keV in some narrow line Seyfert 1 galaxies (Leighly et al. 1997, Dasgupta et al. 2005). First interpreted 
as oxygen absorption in a highly relativistic outflow (v$\sim$0.2-0.3 c), alternative explanations 
have also been proposed since then (Nicastro, Fiore \& Matt, 1999).

\begin{figure}[]
\includegraphics[width=8truecm,height=6truecm,angle=0]{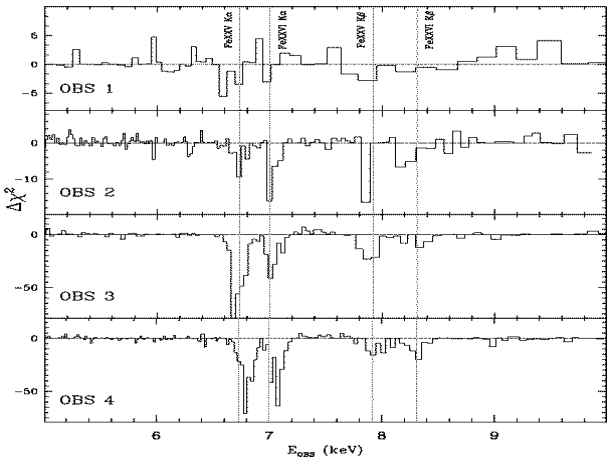}
\caption{ 
$XMM-Newton$ spectral residuals of the Seyfert galaxy NGC1365 by Risaliti et al. (2005). 
Strong (EW $\sim$100 eV each) and 
variable, absorption structures are clearly detected around 7 keV (K$\alpha$ of FeXXV/XXVI) and 8 keV (K$\beta$ of 
 FeXXV/XXVI). Their best-fit absorber model requires outflow velocities of $\sim$5000 km/s, a 
column density of $\sim$5 $\times$ 10$^{23}$ cm$^{-2}$ and a ionization parameter of log $\xi$ = 3.
}
\end{figure}

\begin{figure}[!]
\includegraphics[width=8truecm,height=6truecm,angle=0]{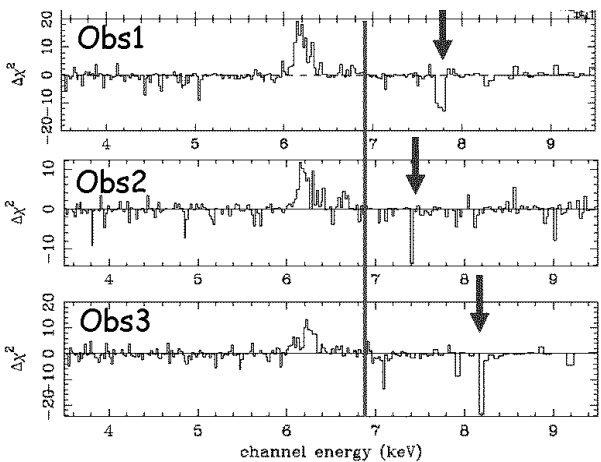}
\caption{
Preliminary residual spectra obtained from multiple $XMM-Newton$ observations of the 
bright Seyfert 1 galaxy Mrk509. Different absorption structures between 7.4 and 8.2 keV are clearly
detected, they are indicated by arrows. The 6.9 keV (rest-frame) energy corresponding to the
H-like Fe ionization level is marked for reference. A strong FeK emission line at 6.2 keV 
(6.4 keV rest-frame) is also clearly visible. Lower time-scale variations of these features 
is under investigation. 
}
\end{figure}

\subsection{Models producing powerfull outflows?}

From a theoretical point of view, there are three main types of models which
are generally called upon to explain AGN winds/outflows: i) thermally driven 
winds, ii) radiatively driven winds and iii) magnetically driven winds.
Thermally driven models are generally most effective in producing slow winds and at large radii. 
Indeed the wind will arise when the gas sound speed is greater than the escape velocity, 
which decreases with radius. They have, for example, been called upon to explain the (large scale) 
warm reflector/absorber seen in the polarized light of Seyfert 2 galaxies 
(Miller and Antonucci 1983, Krolik \& Begelman 1986, 1988; Balsara \& Krolik 1993, 
Krolik \& Kriss 1995, Woods et al. 1996). 
Radiatively and magnetically driven models have instead been mostly proposed as the source of dense and fast 
winds from accretion disks. If the absorbing gas is not too highly ionized, winds can
be driven vertically by radiation from the accretion disk and radially by radiation from 
the central source (e.g. Murray et al. 1995, Proga et al. 2000, and ref. therein). 
Alternatively rotating magnetic field lines can centrifugally accelerate gas above and away 
from the accretion disk plane (Blandford \& Payne 1982, Emmering, Blandford \& Shlosman 1992, 
Kato et al. 2004). Hybrid models where clouds are magnetically elevated above the disk, and radiatively 
accelerated radially, are also being proposed and seem to be quite successfull both theoretically 
(e.g. Proga 2003) and observationally (Elvis 2000). 
Most recent 2/3-dimensional, time-dependent, MHD simulations (Kato et al. 2004, Proga 2005) 
are able to account for dense and extremely fast (v few $\times$ 0.1 c, see Figure 3) flows but the large 
mass outflow rates (Section 2.1) may not be easy to accomodate within current models. 
One example of such a flow model by Kato et al. (2004) is shown in Figure 3.

\begin{figure}[!]
\includegraphics[width=7truecm,height=7truecm,angle=0]{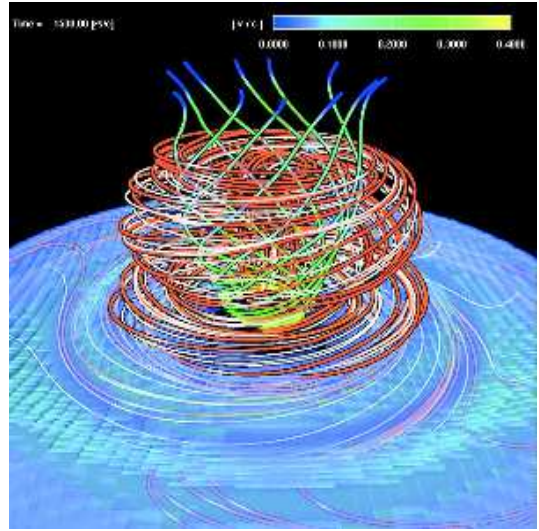}
\caption{
One example of a magnetically-driven accretion flow, by Kato et al. (2004). 
Accumulated toroidal fieldlines emerging from the disk produce a magnetic tower, thereby driving an MHD jet. 
At the base of the flow, velocities as high as 0.4c are produced, as indicated by the color bar.
}
\end{figure}

\subsection{Red-shifted Lines}

Following upon the seminal work by Nandra et al. (1999) on NGC3516, several 
results are now reporting on the detections of absorption structures at rest-frame 
energies between $\sim$4--6 keV in the spectra of several RQ AGNs and QSOs. These have been found
in both Seyfert and RQ QSOs (see Table 1). The two most significant detections to date, 
PG1211+143 and Mrk509, are shown in Figures 4 and 5, respectively.
As for the blueshifted absorption lines illustrated above (Section 2.1), the most natural 
explanation for such absorption components is in terms of resonant absorption 
by highly ionized (H-like or He-like) iron. In this case however the measured energy shifts 
correspond to {\it receding} velocities v$\sim$0.1-0.4 c. 
It is unlikely that the lines result from a lower ionization state than Fe XXV, as in this case a series of 
strong L-shell Fe lines and edges would have been detected at lower energies.
Furthermore, the absorption lines cannot be due to near-neutral Fe (Fe I--XVII) at 6.4 keV, because no 
strong K$\alpha$ absorption line would be observed, as the L shell would be fully populated.

\begin{figure}[]
\includegraphics[width=8truecm,height=6truecm,angle=0]{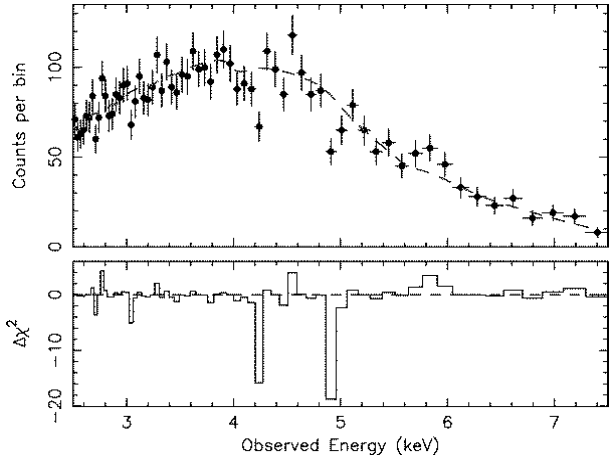}
\caption{
Chandra LETG spectrum of PG 1211+143 (upper panel) and residuals (lower-panel) (Reeves et al. 2005).
Two absorption features are apparent at 4.22 and 4.93 keV, corresponding to 4.56 and 5.33 keV 
in the rest frame of PG 1211+143 (z=0.0809).}
\end{figure}

\begin{figure}[]
\includegraphics[width=8truecm,height=6truecm,angle=0]{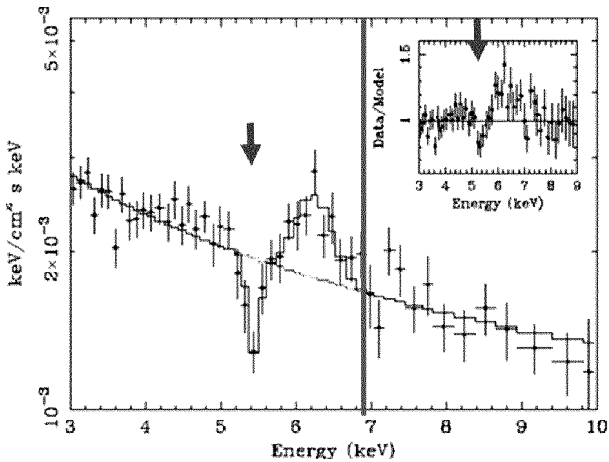}
\caption{
$BeppoSAX$ unfolded spectrum of Mrk509 adapted from Dadina et al. (2005), showing the absorption 
structure (indicated by the arrow) redward of the FeK {\it emission} line visible at $\sim$6.2 keV. 
The energy at 6.9 keV corresponding to H-like Fe is marked by a vertical line. 
Interestingly the absorption-plus-line 
profile is reminiscent of an {\it inverted} P-Cygni profile that would be expected from an 
absorbing {\it inflowing} wind (Longinotti et al. 2006).
{\it Insert:} data-to-model ratios to a simple power-law model. 
}
\end{figure}

\begin{figure}[]
\includegraphics[width=8truecm,height=6truecm,angle=0]{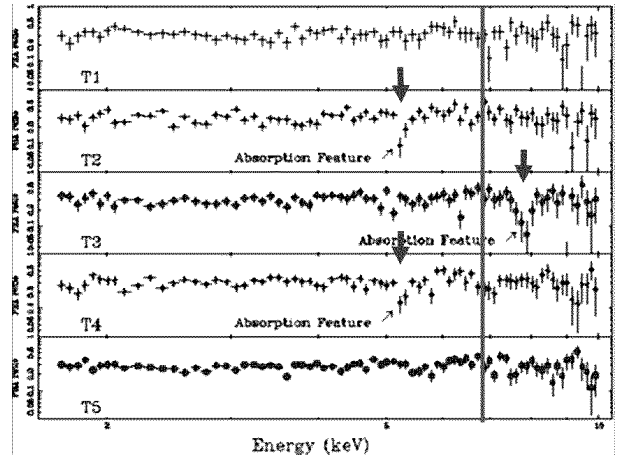}
\caption{
Mrk 509/3C273 PHA ratios during 5 different ($\sim$20 ks) time intervals of 
a ($\sim$100 ks) BeppoSAX observation. Three different absorption features (marked by arrows) 
are clearly visible at E$\sim$5.3 keV (T2 and T4) and 7.9 keV (T3). The features are 
respectively redshifted and blueshifted from the H-like FeK line energy (vertical 
green line at a rest-frame value of 6.9 keV) (see Dadina et al. 2005 for more details).
}
\end{figure}

Modeling of the absorption-line systems with theoretical models such as XSTAR (Kallman et al. 1996) or 
SIABS (Kinkhabwala et al. 2003) require large column densities (typically greater than 10$^{23}$cm$^{-2}$),
very large ionization parameters (greater than $\sim$1000), and mass flow rates of the order of a 
fraction of M$_{\odot}$yr$^{-1}$ (e.g. Dadina et al. 2005, Reeves et al. 2005). These should be 
considered order-of-magnitude estimates due to the uncertainties in the clouds distance from the source, in the 
filling factor and in the black hole mass.

Particularly interesting cases are those of Mrk509 and Mrk335 where the redshifted absorption 
line is imprinted on top of a FeK emission line, thereby forming a kind 
of {\it inverted} P Cygni profile. This is precisely what would be expected (Longinotti et al. 2006) 
from an inflowing gas shell 
which extends over a limited range of radii at a few tens of gravitational radii, R$_g$ 
(R$_g$$\equiv$ GM/c$^2$).
Finally, it should be noted that in Mrk509, red and blue absorption lines have been 
contemporaneously detected, and appear to be transient on time-scales $\lsimeq$20 ks 
(Dadina et al. 2005). Also PG1211+143 has exhibited both type of structures, although not simultaneously, 
suggesting maybe a fundamental link between the two phenomena.  

\subsection{Models producing strong redshift?}

Two possible causes for those (mildly) relativistic redshifted lines have been proposed: either gravitational 
redshift of photons near the massive black hole or infall of matter onto the black hole. 
In the former scenario, the absorbing matter would have to be located very near the black hole, at a few R$_g$.
It could be in the form of a rotating absorbing corona, located on top of the source-plus-disk 
system, as proposed by Ruszkowski \& Fabian (2000), or a kind of ``self-absorption'' effect 
at the inner edge of a nuclear outflow/jet (Yaqoob \& Serlemitsos 2005).  

Alternatively, the matter may be infalling onto the black hole with a 
diffuse (wind) or clumpy (blobs) 
structure. The fact that both Mrk509 and PG1211+143 do also possess 
fast outflows (Section 2.1) 
suggests that inflows may be linked to outflows. One possibility is that 
part of the outflow does not 
escape the gravitational potential of the black hole. For instance in PG1211+143 the material launched with 
v = 0.1c from a radius of R $<$ 100 R$_g$ would not escape the system. This would naturally produce both 
red- and blueshifted lines. A similar situation is predicted by several theoretical models. 
The ``failed disk wind'' model proposed by Proga (2005, see Figure 7), 
the ``thundercloud'' model (Merloni \& 
Fabian 2001), and the ``aborted jet'' model by Ghisellini, Haardt \& Matt (2004) all predict 
episodes of matter ejecta and infalls. Other recent models such as that proposed by Gierlinski \& Done 
(2004) to explain the soft X-ray excess of type-1 AGNs do invoke extreme versions of the 
``failed disk wind'' model, where the absorption structures suffer a huge (v$\sim$0.2-0.3c) 
velocity smearing that is characteristic of a complex velocity structure in the flow (Proga \& Kallman 2004).

The current indications that the red-shifted lines are transient on relatively short ($\sim$10000 s) 
time-scales (Dadina et al. 2005, Reeves et al. 2005, Yaqoob \& Serlemistos 2005) favor 
absorbing systems made of a single dense clump of matter or an ensemble of clumps, rather than a 
diffuse (i.e. wind or corona) structure.
It also calls for the interesting possibility to track the infall of one or more blobs toward the 
black hole, i.e. dispose of ``test-particles'' to probe general relativity (GR) in strong field. 
Regardless of the mechanism, the large redshifts involved, the high covering factor and the sporadic nature of 
the lines imply that the absorber is most likely located within a few R$_g$ of the black hole.

\begin{figure}[]
\includegraphics[width=8truecm,height=6truecm,angle=0]{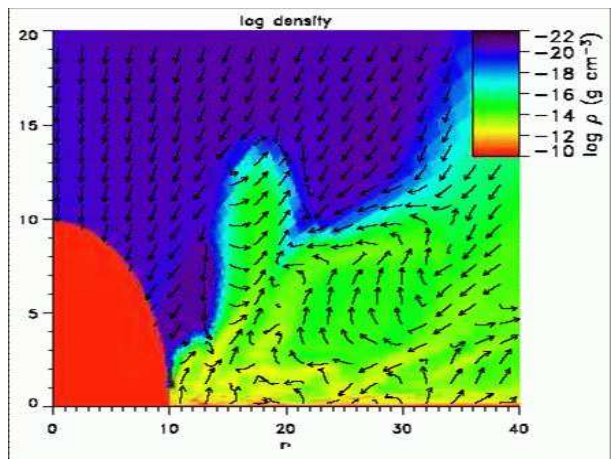}
\caption{
Map of logarithmic density of the AGN failed disk wind by Proga (2005). Note that  
high density (10$^{-12}$-10$^{-14}$ g cm$^{-3}$) flows can develop vertically, and with large 
(up to few 10000 km s$^{-1}$) velocity radial components both inward and outward.
Allowing for a smaller inner radius than used in this computational domain, 
the velocities amplitudes would likely become significantly higher (Proga, private 
communication). 
}
\end{figure}


\section{Critical Issues}


The calculation of the statistical significance of these absorption lines 
is often a matter of debate. Indeed, both blue-shifted and red-shifted absorption 
lines appear to be transient and are detected at energies which are shifted with respect to 
expected atomic values. The analysis thus requires a ``blind'' search in the time and energy 
parameter space. Therefore, statistical significance (and $\chi$$^2$ results) must either correct for the number 
of trials performed or account on proper Monte-Carlo simulations
(i.e. Protassov et al. 2002, Porquet et al. 2004, Yaqoob \& Serlemitsos 2005, Reeves et al. 2005). The statistical 
significance of (single) blue-shifted lines is, to date, robust (typically $\sim$3--5 
$\sigma$), while it is weaker for red-shifted lines (typically $\sim$2--4 $\sigma$). 

Another issue is the difficulty to distinguish, sometimes, between emission and absorption lines. 
For instance, Turner et al. (2002) have shown that the claimed redshifted Fe K-shell absorption in 
the ASCA spectrum of the Seyfert 1 galaxy NGC 3516 (Nandra et al. 1999) was probably due to 
a strong narrow K$\alpha$ core at 6.4 keV and variable emission lines at E$<$6.4 keV which 
were better visible only with the superior {\it XMM-Newton} sensitivity. Similar difficulties 
were encountered by Longinotti et al. (2003) and Dadina \& Cappi (2004) in fitting their data.

The analysis could also be complicated by the difficulty to unambiguously identify (i.e. model) absorption 
lines in the FeK shell energy band (see Kaspi et al. 2006 and Pounds \& Page 2006 in the case of PG1211+134). 
Indeed, (photoionization) modelling of FeK absorption spectra depend sensitively on details of 
energy levels, transition probabilities, and photoionization cross sections 
(i.e. Kallman et al. 2004), hereby potentially affecting the astrophysical interpretation of 
absorption features in the 7--9 keV energy band.

Finally, it has also been claimed that some of these outflows may not exist if absorption features 
that are really local to our Galaxy have been misidentified as being at the redshift of the AGNs 
(McKernan et al. 2004, 2005). This is a caveat that should be beared in mind, in particular 
for sources located near the Galactic plane, though certainly not the case for most of the sources discussed 
here. 


\section{Future Prospects}

The question arises as to whether we are actually seeing only the ``tip of the iceberg''.
Despite red- and blue-shifted FeK absorption lines are of indisputable importance and interest, 
there has been, unfortunately, up to now a number of biases against their detections. 
First there is an ``observational bias'' against the highest-velocity blueshifted features 
(at E$>$7 keV) in that orbiting X-ray telescopes 
are of limited spectral and sensitivity capabilities at these energies. Then the apparently sporadic 
nature of the features, inherent to this type of extreme phenomena, has also generated another bias,  
a ``detection bias''. 
Finally diagnostic of gas with the greatest velocities and, thus, ionization properties will ever only 
be possible through its FeK lines emission/absorption (the lower-Z elements being too weak and/or 
completely ionized), i.e. somewhat a ``physical bias'' against its detectability. 

This, combined to the fact that transient blue- and redshifted absorption lines are naturally 
expected in several theoretical models (Section 2.4) suggests that the parameter space for 
new discoveries is large and important under this topic. In particular, two type 
of measurements are addressed below which may bring in the future to relevant breakthroughs 
in this area: to track the flow dynamics on very short time-scales, and to probe the highest 
velocity outflowing gas.  

\subsection{Gas Dynamics (not only kinematics) around Black Holes}

What are the frequency and duty cycle of this phenomena, what are the typical densities, velocities, 
covering factors and ionization states involved? These are still open issues. 
Despite our lack of detailed understanding, the case is settled for the study of gas inflows and 
outflows in AGNs to give insights into the accretion and ejection processes at work 
around black hole systems. They may offer unprecedented ways of probing/mapping GR under 
the strong field limit and help understand how earth-like quantities of matter (as estimated in 
Dadina et al. 2005) are accelerated up to relativistic velocities and kept collimated. 
The most important goals would thus be to characterise the geometry, the kinematics (measure 
of velocities, v) and, most importantly, the dynamics (measure of velocity variations, 
$\Delta$v/$\Delta$t) of the absorbing material, either 
inflowing or outflowing. For example, one wish to track the absorption lines from, say, $\sim$1 Rs(=2 Rg) to 10 Rs, 
with intervals of $\sim$1 Rs. Assuming a velocity of v$\sim$0.2 c, these correspond to time-scale intervals of 
$\sim$5000 s for a black hole of mass 10$^8$ M$\odot$ and $\sim$ 50 s for a 10$^6$ M$\odot$ black hole.
Assuming an equivalent width of $\sim$ $-$100 eV for the Fe absorption line and a bright source 
with 2-10 keV flux of 2$\times$10$^{-11}$ erg cm$^{-2}$s$^{-1}$ (as measured in Mrk509, 
Dadina et al. 2005), simulations show that XEUS could detect such line at any
energy between 4--10 keV and on a time-scale as short as 100 s.
Inflowing, outflowing, accelerating and/or decelarating absorption lines could then be 
tracked (as illustrated in Figure 8) on few tens of AGNs on a few 100 s time-scale. 
Clearly, very high throughput between 4--10 keV is mandatory for this type of measurements.

\begin{figure}[]
\includegraphics[width=8truecm,height=6truecm,angle=0]{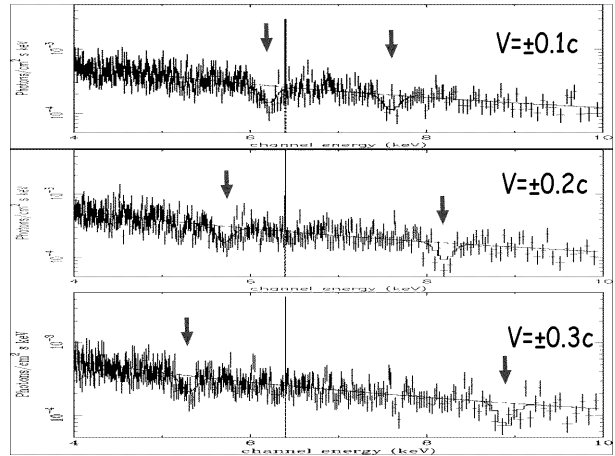}
\caption{ 
Simulated 100 s XEUS (with 2m$^{2}@$6 keV) observations of a ``Mrk509-like'' source to highlight 
its capabilities to study the flow dynamics (i.e. track velocity variations). 
Two absorption lines expected from highly 
ionized blob(s) inflowing and/or outflowing at v=0.1, 0.2, and 0.3c are simulated 
(from top to bottom). The EWs used 
are -100 eV, all data points are rebinned to S/N$\gsimeq$3 and a 6.4 keV energy is 
marked with a vertical line for reference. 
}
\end{figure}

\subsection{Highest Velocity Outflows as Source of Cosmological Feedback from AGNs}

It is well established in litterature that 'slow' (non-relativistic) AGN winds may have an important cosmological 
impact in that they may provide an important source of energy feedback (see the review by Elvis 
2006 and references therein). Their effect on a number 
of astrophysical areas (such as the AGN-host galaxies co-evolution, the heating of galaxies, 
groups and cluster of galaxies, the disruption of cooling flows, the ISM and IGM 
metal enrichment, etc.) is not yet established because of 
the remaining large (order of magnitudes\footnote{Current estimates for the mass outflow rates 
of AGN winds span values from a few \% to several times the AGN mass accretion rates.}) 
uncertainties in the total mass, energy and momentum 
released into the ISM and IGM (Blustin et al. 2005, Creenshaw et al. 2003, Chartas et al. 2004, Elvis 2006,).
Likely, their impact would be even more relevant if extreme outflows like the ones presented here 
are considered (Pounds et al. 2003, King \& Pounds 2003). The remarkable detections in at least a few 
high-z BAL QSOs (Chartas et al. 2003) have then set the ground for quasar massive outflows to have a very 
significant impact in the evolution of their own host galaxies (Chartas, Brandt \& Gallagher 2004;
Lapi, Cavaliere \& Menci 2005). How ubiquitous are these massive outflows in high-z BAL and non-BAL 
QSOs has yet to be understood and will necessitate further observations. 

As stated in Section 4, we may have been missing most of the highest-velocity gas because of  
a number of biases against its detection.  
These are precisely the outflows that are most interesting and potentially most important because 
they could transport outward most of the mechanical energy emerging from the black holes.
Good sensitivity and energy resolution are needed at all energies from $\sim$7 up to $\sim$15 keV 
to be able to detect and constrain absorption lines and edges from the highest ionization Fe (up to 9 
keV rest-frame), at the highest plausible velocities (up to, say, 0.9c). 
Much improvement with respect to XMM-Newton will be obtained with the planned franco-italian 
mission Simbol-X (see simulations in Figure 9). More detailed, time-resolved, spectroscopy 
on shorter time-scales and/or fainter sources shall wait the hard X-ray detector planned on-board 
XEUS (see ESA's XAWG Report: Parmar et al. 2004). 

\begin{figure}[]
\includegraphics[width=8truecm,height=6truecm,angle=0]{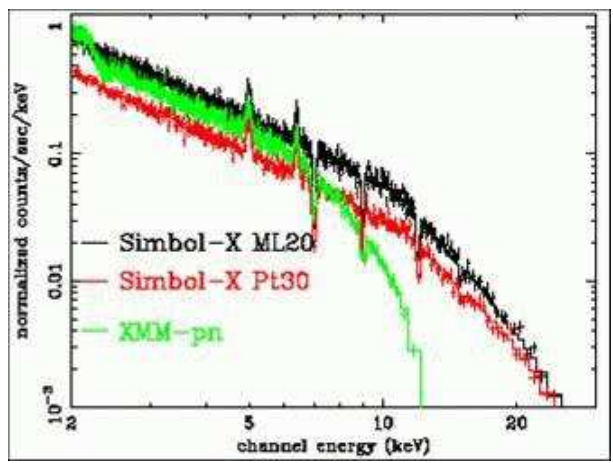}
\caption{
Simulations of a 50 ks observation with Simbol-X (dark and red curves for two different mirror 
configurations), compared to a simulated 50 ks XMM-Newton observation (green curve). 
The model consists in a power-law with $\Gamma$=1.9 and F(2-10)=10$^{-11}$ erg cm$^{-2}$ s$^{-1}$, plus 
2 narrow emission lines (at 5 and 6.4 keV, with EW=100 eV), plus 4 
absorption lines at 7, 9, 12, and 15 keV ($\sigma$ $<$50 eV, EW=-100 eV). 
The XMM-Newton simulations would (barely) detect the two absorption lines below 10 keV, while 
Simbol-X would detect them all, up to $\sim$15 keV.
}
\end{figure}

\section{Conclusion}

Recent {\it Chandra} and {\it XMM-Newton} observations of bright Seyfert galaxies and 
QSOs have revealed the presence of massive outflows of ionized material driven from 
a few gravitational radii from the black holes with velocities up to $\sim$0.4c.
Other recent observations show evidence for redshifted iron K$\alpha$ absorption lines, 
which require either pure gravitational redshift or matter infalling directly onto the black hole. 
The two phenomena (relativistic outflow and inflow) may possibly be linked together 
and there are observational and theoretical evidence for
this to be true. Both phenomena are of outstanding interest because they offer new potential to probe 
the dynamics of innermost regions of accretion flows, to tackle the formation of ejecta/jets and 
to place constraints on the rate of kinetic energy injected by AGNs into the ISM and IGM.  
Future high energy missions (such as the planned Simbol-X and XEUS) will likely allow an exciting step 
forward in our understanding of the flow dynamics around black holes and the formation of 
the highest velocity outflows. 

\acknowledgements

I gratefully thank Stefano Bianchi, Mauro Dadina, Giorgio Palumbo, Gabriele Ponti 
and Cristian Vignali for a carefull reading and commenting on an earlier version of this 
manuscript.
Based on observations obtained with XMM-Newton, an ESA science mission with instruments and 
contributions directly funded by ESA Member States and NASA. I acknowledge financial support 
from ASI under contract I/023/05/0.


{}

\end{document}